\documentclass[10pt,letterpaper,american]{article}
\usepackage[latin9]{inputenc}
\usepackage{color}
\usepackage{babel}
\usepackage{booktabs}
\usepackage{amsmath}
\usepackage{amssymb}
\usepackage{graphicx}
\usepackage[unicode=true,
 bookmarks=false,
 breaklinks=false,pdfborder={0 0 1},backref=false,colorlinks=true]
 {hyperref}
\hypersetup{
 citecolor=blue,urlcolor=blue}

\makeatletter

\pdfpageheight\paperheight
\pdfpagewidth\paperwidth

\providecommand{\tabularnewline}{\\}

\usepackage{osameet3}
\usepackage{capt-of}
\usepackage{booktabs}
\usepackage[table]{xcolor}

\usepackage{cite}
\usepackage[printonlyused]{acronym}
\acrodef{DM}{distribution matcher}
\acrodef{Hi-DM}{hierarchical DM}
\acrodef{PS}{probabilistic shaping}
\acrodef{PAS}{probabilistic amplitude shaping}
\acrodef{AWGN}{additive white Gaussian noise}
\acrodef{QAM}{quadrature amplitude modulation}
\acrodef{FEC}{forward error correction}
\acrodef{PAM}{pulse amplitude modulation}
\acrodef{LUT}{look up table}
\acrodef{MB}{Maxwell-Boltzmann}
\acrodef{ESS}{enumerative sphere shaping}
\acrodef{CCDM}{constant composition distribution matching}
\acrodef{IR}{information rate}
\acrodef{SNR}{signal to noise ratio}
\acrodef{PPM}{pulse position modulation}
\acrodef{invDM}{inverse DM}
\acrodef{TX}{transmitter}
\acrodef{RX}{receiver}
\acrodef{BER}{bit error rate}
\acrodef{SER}{symbol error rate}

\makeatother

\begin{document}

\title{Coupled-Channel Enhanced SSFM for Digital Backpropagation in WDM
Systems}

\author{S. Civelli$^{1,2}${*}, E. Forestieri$^{1,2}$, A. Lotsmanov$^{3}$,
D. Razdoburdin$^{3,4}$, M. Secondini$^{1,2}$}

\address{$^{1}$ TeCIP Institute, Scuola Superiore Sant'Anna, Via G. Moruzzi
1, 56124, Pisa, Italy\\
$^{2}$ PNTLab, CNIT, Via G. Moruzzi 1, 56124, Pisa, Italy\\
$^{3}$ Moscow Research Center, Huawei Technologies Co., Ltd., Moscow,
Russia \\
$^{4}$ Sternberg Astronomical Institute, Moscow M.V. Lomonosov State
University, Moscow, Russia}

\email{{*}stella.civelli@santannapisa.it}

\maketitle
\copyrightyear{2021}
\begin{abstract}
A novel technique for digital backpropagation (DBP) in wavelength-division
multiplexing systems is introduced and shown, by simulations, to outperform
existing DBP techniques for approximately the same complexity. 
\end{abstract}


\section{Introduction}

Digital backpropagation (DBP) is widely investigated for the mitigation
of nonlinearity in optical fiber communication, with the split step
Fourier method (SSFM) being the most popular method to practically
implement DBP \cite{Essiambre:ECOC05,kahn_bp}. The SSFM digitally
inverts the optical channel by applying a number of steps in which
linear and nonlinear effects are considered separately. Several alternative
DBP techniques have been proposed to achieve the same performance
as the SSFM with less steps and, hence, lower complexity, such as
the correlated or filtered DBP \cite{Li:OFC11,Rafique:OE2011} and
the enhanced SSFM (ESSFM) \cite{secondini2014enhanced,secondini_PNET2016}.

While both the ESSFM and the classical SSFM are conceived for the
propagation of a single channel, they can also be used for several
wavelength-division multiplexing (WDM) channels, as long as they are
jointly represented as a single optical field---we refer to this
as \emph{full-field} DBP. The potential gain achievable by full-field
DBP increases with the number of jointly backpropagated channels \cite{dar_JLT2017_nonlinear}.
However, the extremely large number of steps required in this case
makes the practical implementation of full-field DBP not feasible
\cite{liga2014performance}.

In this work, we propose a novel technique for DBP---the\emph{ coupled-channel}
ESSFM (CC-ESSFM)---which improves the ESSFM by explicitly accounting
for the cross phase modulation (XPM) generated by several copropagating
channels and for its interplay with dispersion. We show, through simulations,
that the CC-ESSFM achieves better performance than single- and full-field
SSFM and ESSFM techniques for the same complexity.

\section{Coupled-channel enhanced SSFM\label{sec:CCESSFM}}

\begin{figure}
\begin{centering}
\includegraphics[width=1\textwidth]{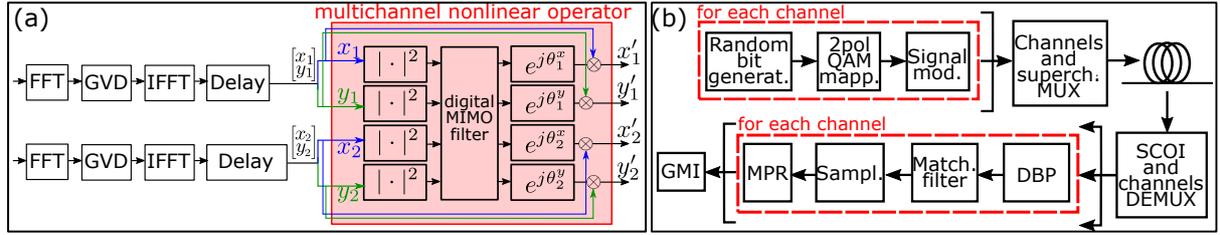}\vspace*{-1ex}
\par\end{centering}
\caption{\label{fig:ESSFM-multichannel}(a) Linear and nonlinear step of the
CC-ESSFM for $N_{\mathrm{ch}}=2$ channels. (b) System setup}
\vspace*{-2ex}
\end{figure}

The CC-ESSFM is described below for $N_{ch}$ copropagating channels.
The CC-ESSFM, like the SSFM and the ESSFM, consists of several steps
$N_{s}$ made of a linear and a nonlinear part, and is depicted in
Fig.~\ref{fig:ESSFM-multichannel}(a).

In the linear part, the various channels are independently processed
as in the SSFM or ESSFM, with the only difference that a time delay
is introduced to account for the different group velocity of the channels
and ensure that their contributions are correctly synchronized at
each nonlinear step.

In the nonlinear part, a nonlinear phase rotation is applied to each
polarization component of each channel to account for the self-phase
modulation (SPM) and part of the XPM. According to the frequency-resolved
logarithmic perturbation (FRLP) model \cite{Secondini:JLT2013-AIR},
each channel induces a frequency-dependent nonlinear phase rotation
which can be expressed as a quadratic form of its samples and affects
the channel itself (SPM) and the other channels (XPM). The ESSFM accounts
only for the SPM contribution and approximates it by a filtered version
of the signal intensity, where the filter coefficients are numerically
optimized to minimize the error \cite{secondini2014enhanced,secondini_PNET2016}.
In the CC-ESSFM, on the other hand, also the XPM contributions are
accounted for and expressed as filtered versions of the corresponding
signal intensities. By assuming that each channel is separately represented
over its bandwidth with $n$ samples per symbol, the nonlinear step
for the $i$th channel can be expressed as $x_{i}'[k]=x_{i}[k]\exp(j\theta_{i}^{x}[k])$
and $y_{i}'[k]=y_{i}[k]\exp(j\theta_{i}^{y}[k])$, where $x_{i}[k]$
and $y_{i}[k]$ (and $x_{i}'[k]$ and $y_{i}'[k]$) are the $k$th
samples of the two polarization components of the $i$th channel at
the input (and output) of the nonlinear step, normalized to have unit
power;
\begin{align}
\theta_{i}^{x}[k] & =-\sum_{\ell=1}^{N_{ch}}\left(\overline{\ensuremath{\phi}}_{i\ell}^{\parallel}\sum_{m=-N_{c}}^{N_{c}}C_{\ell-i}[m]|x_{\ell}[k+m]|^{2}+\overline{\ensuremath{\phi}}_{\ell}^{\perp}\sum_{m=-N_{c}}^{N_{c}}C_{\ell-i}[m]|y_{\ell}[k+m]|^{2}\right)\label{eq:nonlinear_phase_x}\\
\theta_{i}^{y}[k] & =-\sum_{\ell=1}^{N_{ch}}\left(\overline{\ensuremath{\phi}}_{\ell}^{\perp}\sum_{m=-N_{c}}^{N_{c}}C_{\ell-i}[m]|x_{\ell}[k+m]|^{2}+\overline{\ensuremath{\phi}}_{i\ell}^{\parallel}\sum_{m=-N_{c}}^{N_{c}}C_{\ell-i}[m]|y_{\ell}[k+m]|^{2}\right)\label{eq:nonlinear_phase_y}
\end{align}
are the (different) nonlinear phase rotations over the two polarization
components; $\overline{\ensuremath{\phi}}_{\ell}^{\perp}=\gamma P_{\ell}\int_{z-L/2}^{z+L/2}g(z)dz$
and $\overline{\ensuremath{\phi}}_{i\ell}^{\parallel}=(2-\delta_{i\ell})\overline{\ensuremath{\phi}}_{\ell}^{\perp}$
are the average nonlinear phase rotations induced, respectively, by
the copolarized and orthogonal component of the $\ell$th channel
over the $i$th channel; $\delta_{i\ell}$ is the Kronecker delta
($\delta_{i,\ell}=1$ if $i=\ell$ and $\delta_{i,\ell}=0$ otherwise);
$P_{\ell}(z)$ is the power profile of the $\ell$th channel along
the step of length $L$; $\gamma$ is the nonlinear coefficient; and
$C_{h=\ell-i}[m]$, with $m=-N_{c},\ldots,N_{c}$, are the $2N_{c}+1$
real ESSFM coefficients which account for the effect of dispersion
on the XPM generated by the $\ell$th channel over the $i$th channel.

With respect to the standard SSFM, the CC-ESSFM requires, at each
step, the implementation of a digital $2N_{ch}\times2N_{ch}$ multi-input
multi-output (MIMO) filter for the computation of (\ref{eq:nonlinear_phase_x})
and (\ref{eq:nonlinear_phase_y}). This can be efficiently done in
frequency domain by using a pair of real FFTs and the corresponding
IFFTs per each channel, with an overall increase in complexity of
less than 50\% compared to the SSFM. On the other hand, the single-channel
ESSFM---equivalent to the CC-ESSFM with $N_{\mathrm{ch}}=1$---is
more efficiently implemented in time-domain when the number of coefficients
$N_{c}$ is not too large. Table~\ref{tab:complexity-short} reports
the complexity of the ESSFM and CC-ESSFM compared to the standard
SSFM and a simple dispersion compensation, considering the same implementation
assumptions as in \cite{secondini_PNET2016} and an overlap-and-save
processing, where $N$ is the ``overlapping'' block length and $\eta$
the ratio between $N$ and the actual number of samples ``saved''
per each processed block.

The CC-ESSFM coefficients can be obtained either by numerical optimization
or analytically. In this work, we have adopted the numerical optimization,
which guarantees a good performance and does not require the a priori
knowledge of the link parameters, deferring the study of the analytical
computation based on the FRLP model \cite{Secondini:JLT2013-AIR}
to a future work. The numerical optimization is performed by assuming
that the coefficients are independent of the launch power and satisfy
the symmetry condition $C_{h}[m]=C_{-h}[-m]$, as predicted by the
FRLP model and verified by numerical simulations. To further simplify
the optimization, the coefficients $\mathbf{C}_{h}=(C_{h}[-N_{c}],\ldots C_{h}[N_{c}])$
are obtained iteratively for $h=0,\dots,N_{ch}-1$, assuming $N_{c}=32$
for $h=0$ and $N_{c}=128$ for $h>0$. At the $h$th iteration, $\mathbf{C}_{h}$
is obtained by minimizing the mean square error over the received
symbols, while keeping $\mathbf{C}_{0},\dots,\mathbf{C}_{h-1}$ fixed
to the values found at the previous iterations, and setting $\mathbf{C}_{h+1},\dots,\mathbf{C}_{N_{\mathrm{ch}}-1}$
to zero.

\section{System setup and performance\label{sec:systemsetupperformance}}

\begin{figure}
\begin{minipage}[b][1\totalheight][t]{0.66\columnwidth}%
\makeatletter
\long\def\@makecaption#1#2{%
  \normalsize
  \vskip\abovecaptionskip
  \parbox{3in}{#1. #2}
  \vskip\belowcaptionskip\normalsize}
\makeatother\centering

\includegraphics[width=1\textwidth]{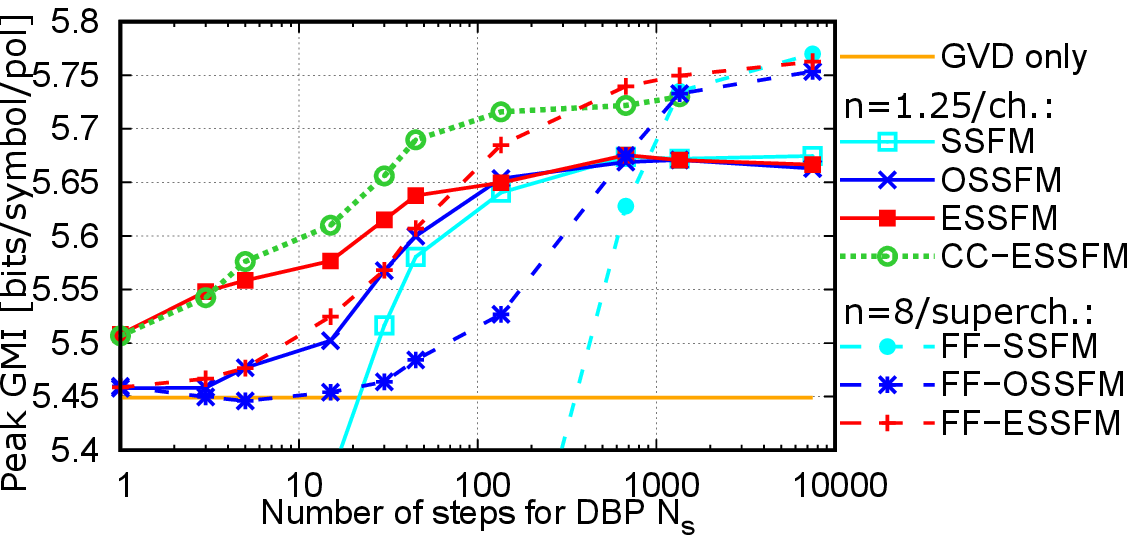}

\captionof{figure}{\label{fig:MinBER_noBPS_U_FF}Performance of different
DBP techniques versus number of steps $N_{s}$.}%
\end{minipage}\hfill{}%
\begin{minipage}[b][1\totalheight][t]{0.33\columnwidth}%
\makeatletter
\long\def\@makecaption#1#2{%
  \normalsize
  \vskip\abovecaptionskip
  \parbox{2.in}{#1. #2}
  \vskip\belowcaptionskip\normalsize}
\makeatother\centering\renewcommand{\arraystretch}{2.05}\setlength{\tabcolsep}{1pt}\rowcolors{2}{}{gray!10}

\begin{tabular}{lr}
\toprule 
Technique & Real multipl./4D~symb.\tabularnewline
\midrule
GVD only & $\eta n(8\log_{2}N{+}8)$\tabularnewline
SSFM & $N_{s}\eta n(8\log_{2}N{+}21)$\tabularnewline
ESSFM & $N_{s}\eta n(8\log_{2}N{+}21{+}N_{c})$\tabularnewline
CC-ESSFM & $\begin{gathered}N_{s}\eta n\bigl(12\log_{2}N\quad\\
\qquad\qquad{+}20{+}4N_{\mathrm{ch}}\bigr)
\end{gathered}
$\tabularnewline
\bottomrule
\end{tabular}

\captionof{table}{\label{tab:complexity-short}Number of real multiplications
per 4D symbol per channel.}%
\end{minipage}\vspace*{-1ex}
\end{figure}
The simulation setup is sketched in Fig.\ \ref{fig:ESSFM-multichannel}(b).
The superchannel of interest (SCOI) is made of $N_{ch}=4$ channels
with baud rate $R_{s}=41.67$GBd, $75$GHz spacing, $64$ quadrature
amplitude modulation (QAM), and a root-raised-cosine pulse-shaping
filter with a roll-off of 0.1. Two side superchannels with the same
characteristics of the SCOI are included in the simulations. The link
is made of $15$ spans of $80$~km single mode fiber (SMF) ($D=17$ps/nm/km,
$\gamma=1.3$/W/km, $\alpha_{\text{dB}}=0.2$dB/km); after each span,
an erbium-doped fiber amplifier (EDFA) with a noise figure of $5$dB
compensates for the loss. At the receiver side, after SCOI and channels
demultiplexing, each channel of the SCOI undergoes DBP---separately,
or jointly when CC-ESSFM is employed---, matched filtering, and symbol-time
sampling. On the other hand, when full-field DBP is used, it is performed
over the entire SCOI, before channel demultiplexing. Finally, after
removing a possible residual mean phase rotation (MPR) caused by nonlinearity
and not compensated for by DBP, the average generalized mutual information
(GMI) over the $4$ channels of the SCOI is evaluated \cite{alvarado2015replacing}.
The number of samples per symbols considered for DBP is $n=1.25$
(per channel) for single-field and coupled-channel DBP, and $n=8$
(per superchannel) for full-field DBP. The optimization of the coefficients
 and the evaluation of the GMI are performed using a stream of $1024$
and $65536$ symbols per channel, respectively.

Figure~\ref{fig:MinBER_noBPS_U_FF} compares the peak GMI obtained
at optimal launch power with different DBP techniques as a function
of the number of steps $N_{s}$. For a small (but practical) number
of steps, the conventional SSFM has a very poor performance---in
fact, even worse than simple GVD compensation. In this case, the performance
can be improved (without changing the complexity) by using the optimized
SSFM (OSSFM), with a numerically optimized nonlinear coefficient (lower
than the actual one) in the nonlinear step \cite{liga2014performance}.
An even better performance is achieved by the ESSFM, while the best
performance is achieved by the CC-ESSFM. In this practical scenario,
the full-field techniques (dashed lines) perform significantly worse
than the corresponding single-channel ones (solid lines), due to the
larger accumulated dispersion in each step. The picture changes when
the number of steps is increased to less realistic values. In this
case, the full-field techniques outperform the corresponding single-channel
techniques and perform even slightly better than the CC-ESSFM. Asymptotically
($N_{s}\rightarrow\infty$), all the single-channel techniques converge
to the same limit, corresponding to an exact compensation of intrachannel
effects. Analogously, all the full-field techniques converge to a
higher limit, corresponding to an exact compensation of intra-superchannel
effects, while  CC-ESSFM converges to a slightly lower limit, as it
accounts only for SPM and XPM in the SCOI.

Though the number of steps is a good indicator of the relative complexity
of the various algorithms, the actual complexity depends on the specific
algorithm and on the details of the implementation. Table~\ref{tab:complexity-short},
obtained under some reasonable but simplifying assumptions, allows
a more accurate comparison and shows that ESSFM and CC-ESSFM require,
respectively, about 30\% and 50\% more multiplications per step than
SSFM, hence only slightly reducing their advantage compared to the
other techniques in Fig.~\ref{fig:MinBER_noBPS_U_FF}. However, we
believe that additional simplifications and a careful implementation
can further reduce these figures.

Similar results (not shown here for lack of space) have been obtained
when considering probabilistically-shaped QAM modulations and when
including a practical carrier phase recovery algorithm at the receiver.

\section{Conclusion}

We have proposed the CC-ESSFM as an efficient technique for multi-channel
DBP in WDM systems. By additionally accounting for XPM among the channels
and its interplay with dispersion, the CC-ESSFM outperforms conventional
single-channel and full-field DBP techniques for the same (sufficiently
small) number of steps.

\bibliographystyle{osajnl}
\bibliography{ref,refs_kw}

\end{document}